\documentclass[prd,preprintnumbers,nofootinbib,aps,9pt]{revtex4}
\usepackage{subeqnarray}
\usepackage{epsfig,amsmath,amssymb}
\usepackage{mathrsfs}
\usepackage[usenames,dvipsnames]{color}
\usepackage[pagebackref=true, colorlinks=true]{hyperref}
\definecolor{redish}{rgb}{0.7,0.2,0.0}  
\definecolor{bluish}{rgb}{0.2,0.5,0.8}
\hypersetup{linkcolor=redish,          
                  citecolor=blue,        
                  filecolor=magenta,      
                  urlcolor=bluish}          
\DeclareFontFamily{U}{rsfs}{}         
\DeclareFontShape{U}{rsfs}{m}{n}{<5> rsfs5 <6><7> rsfs7          %
  <8><9><10><10.95><12><14.4><17.28><20.74><24.88> rsfs10}{}     %
\DeclareMathAlphabet{\mathfs}{U}{rsfs}{m}{n}                     %
\newcommand{\mfs}[1]{\mathfs {#1}}                               %

\newcommand{\ba}{\nopagebreak[3]\begin{eqnarray}}
\newcommand{\ea}{\end{eqnarray}}
\newcommand{\bii}{\begin{itemize}}
\newcommand{\eii}{\end{itemize}}
\newcommand{\nn}{\nonumber}

\newcommand{\sO}{{\mfs O}}

\newcommand{\f}{\frac}

\def \d{\delta}

\def \l{\ell}
\def \g{\gamma}

\def \lp{\l_p}
\def \j{\sqrt{j(j+1)}}

\def \lm{\lambda}

\def \s{\sigma}

\def \sj{s_j^{\star}}

\def \({\left(}
\def \){\right)}
\def \[{\left[}
\def \]{\right]}
\begin{document}
\title{Proof of Bekenstein-Mukhanov ansatz in loop quantum gravity}
\author{Abhishek Majhi}%
 \email{abhishek.majhi@gmail.com}
\affiliation{School of Physics, Indian Institute of Science Education and Research\\ Thiruvananthapuram (IISER TVM), Trivandrum 695016, IISER Trivandrum, India\\}
\affiliation{Astro-Particle Physics and Cosmology Division\\Saha Institute of Nuclear Physics\\Kolkata 700064, India}%
\begin{abstract}
A simple proof of Bekenstein-Mukhanov(BM) ansatz is given within the loop quantum gravity(LQG) framework. The macroscopic area of an equilibrium black hole horizon indeed manifests a linear quantization. The quantum number responsible for this discreteness of the macroscopic area has a physical meaning in the LQG framework, unlike the ad hoc one that remained unexplained in BM ansatz.  
\end{abstract}
\maketitle
\section{Introduction}
 In \cite{bm} J. D. Bekenstein and V. F. Mukhanov({\bf BM})  studied the decay of the area of a radiating massive black hole.  Having no quantum theory of black holes at their disposal, BM {\it assumed} that the area {\it spectrum} of a black hole is given by 
\ba
A=\alpha N\lp^2\label{an}
\ea
where $\alpha$ is a pure number and $\lp=\sqrt{G\hbar/c^3}$ is the Planck length and $N$ is some sort of quantum number whose origin and physical meaning remained obscure. The assumption was mainly motivated by the discrete energy spectrum of atoms in ordinary quantum mechanics. Following this ansatz BM came to the conclusion that the radiation from black holes is discontinuous and non-thermal in nature as opposed to the prediction of Hawking\cite{haw}. 

Here, I shall show that simple application of statistical mechanics in the quantum theory of equilibrium black holes in {\it loop quantum gravity}({\bf LQG}) yield that, in spite of the complicated nature of the area spectrum of black hole horizon  at the microscopic level, the macroscopic or classical area of the horizon$(A_{cl})$, manifest a linear discreteness {\it resembling} eq.(\ref{an}) and is given by $A_{cl}=\tilde\alpha N\lp^2$. The quantity $\tilde\alpha$ is obtained, rather implicitly, by demanding that the leading order term of the entropy be given by $A_{cl}/4\ell_p^2$ as observed by an asymptotic observer. Most interestingly, the quantum number $N$ which is responsible for the discreteness of the macroscopic area has a physical meaning and its origin is completely understood in the LQG framework. 

It is worthy of mentioning that the technical parts and the calculations which will be discussed in this work can be found in detail in some earlier works\cite{me1,ampm2,me3}. But the observation made and the interpretation of the result provided here are the important value additions of this work which might be of interest to the concerned reader.

\section{Black hole horizon in LQG }
Modern day description of an equilibrium black hole horizon is captured by the local notion of an {\it isolated horizon}({\bf IH}), which is a generalization of the global concept of event horizon to a more realistic situation where matter and radiation are allowed arbitrarily close to and outside the IH, but not allowed to cross the IH\cite{ih1,ih2}. Technically, an IH is a null, topologically $S^2\times R$, inner boundary of the spacetime satisfying certain boundary conditions so as to capture the properties of an equilibrium black hole horizon from a completely local perspective independent of the dynamics of the bulk off the horizon\cite{ih1,ih2}. The phase space analysis of the IH reveals that there is an $SU(2)$ {\it Chern-Simons}({\bf CS}) theory on the IH\cite{ih1,ih2}. The CS term appears in the action with a coupling constant $k$, called the CS level and is given by $k\equiv A_{cl}/4\pi\g\lp^2$, where $\g$ is the Barbero-Immirzi parameter\cite{bar1,bar2,im1,im2} and $A_{cl}$ is the classical area of a cross-section of the IH.  A {\it quantum isolated horizon}({\bf QIH}) can be visualized as an IH punctured by the edges of the spin network graph which span the bulk quantum geometry in LQG\cite{qg1,qg2}. In the theory of QIH,  $k$ needs to be an integer which is a pre-quantization condition. The physical quantum states of the QIH are that of the CS theory coupled to the punctures which act as sources. The edges of the spin network endow the punctures with the corresponding spins which give rise to the quantum area spectrum of the QIH\cite{qg1,qg2}. 

 \section{ Proof of BM Ansatz}
\subsection{Quantum Statistical Mechanics}
 Let us consider that there are $N$ punctures on the QIH carrying spins $j_1,j_2, \cdots,j_N$ where all the spins can take values $1/2, 1, 3/2, \cdots, k/2$. Alternatively one can consider that $s_j$ punctures carry spin $j$ so that $\sum_{j=1/2}^{k/2}s_j=N$. These spin configurations provide the area eigenstate basis. Such a basis state of the QIH Hilbert space is denoted by the ket $|\{s_j\}\rangle$ having the area eigenvalue given by 
\ba
\hat A|\{s_j\}\rangle=8\pi\g\lp^2\sum_{j=1/2}^{k/2} s_j\j~|\{s_j\}\rangle\label{area}
\ea
Such a spin configuration (eigenstate) has a $(N!/\prod_j s_j!)$-fold degeneracy due to the possible arrangement of the spins yielding the same area eigenvalue. 
A macroscopic state of a QIH is designated by the two integer parameters $k$ and $N$\cite{gp,ampm2} which can be appropriately written as
$
|k,N\rangle=\sum_{\left\{s_j\right\}}c[\left\{s_j\right\}]|\left\{s_j\right\}\rangle\nn
$ 
where $|c[\left\{s_j\right\}]|^2=\omega[\left\{s_j\right\}]$(say) is the probability that the QIH is found in the state $|\left\{s_j\right\}\rangle$.

In \cite{me1}, it has been shown, using  the method of most probable distribution, that the entropy of such a macrostate of the QIH i.e. for given $k$ and $N$, is given by 
\ba
S={\lm A_{cl}}/{8\pi\g\lp^2}+N\s~~,\label{lent}
\ea
ignoring the logarithmic correction with universal coefficient -3/2\cite{ampm2,km00}. It should be mentioned that all the calculations are valid in the limit $k,N\to\infty$ which is physically reasonable for large horizons which are being dealt with in this work. The steps can be debriefed as follows :
\begin{enumerate}
\item Logarithm of the number of microstates $\Omega(k,N)=\sum_{\{s_j\}}\Omega[\{s_j\}]$ is the entropy corresponding to a macrostate $|k,N\rangle$. According to the basic postulates of equilibrium statistical mechanics, one of the spin configurations is the most probable one and the number of microstates of any other spin configuration is negligible compared to the most probable one. Hence, $\Omega(k,N)\simeq\Omega[\{\sj\}]$
\item Maximizing the entropy $\Omega[\{s_j\}]$ corresponding to a spin configuration $\{s_j\}$ subject to the two constraints $C_1 : \sum_j s_j\j=k/2$ and $C_2 : \sum_js_j=N$ yield the most probable spin distribution.
\item The distribution for the dominant spin configuration $\{\sj\}$ i.e. the most probable spin distribution comes out to be 
\ba
\sj=N(2j+1)\exp\left[-\lm\j-\s\right]\label{mpd}					
\ea
where $\lm$ and $\s$ are the two Lagrange multipliers corresponding to the constraints $C_1$ and $C_2$ respectively.
\item Taking the logarithm of $\Omega[\{\sj\}]$, a few steps of algebra along with some approximations in the appropriate limits yield the form of the entropy mentioned above in eq.(\ref{lent}).\footnote{ The details of the entropy calculation is completely irrelevant in the present context and an enthusiast is advised to look into \cite{me1,ampm2} for detailed calculations.} 
\item Satisfying $C_1$ and $C_2$ with $\sj$ and taking the appropriate limits one obtains respectively
\ba
\f{k}{N}&=&1+\f{2}{\lm}+\f{4}{\lm(\sqrt 3\lm+2)}\label{lmro}\\
e^{\s}&=&\f{2}{\lm^2}\left(1+\f{\sqrt 3}{2}\lm\right)e^{-\f{\sqrt 3}{~2}\lm}\label{siglm}
\ea
\end{enumerate}
{\it Remarks :} It may be pointed out that the constraint $C_1$ is an approximate condition which is satisfied only in the limit $s_j, N,k\ggg\sO(1)$. Otherwise it makes no sense because the left hand side of $C_1$ is a weighted sum of irrational quantities, whereas the right hand side is a rational number. The constraint $C_1$ dictates the fact that we are looking at the area eigenstates within $\pm\sO(\lp^2)$ window about the classical area $A_{cl}$. From mathematical viewpoint, these limits allow us to apply the usual methods of equilibrium statistical mechanics to this scenario which resembles(but not equivalent to) usual gas thermodynamics. From the physical viewpoint, these limits are the appropriate ones for the quantum description of IH. The reasons can be explained as follows. Firstly, if the horizon area becomes small enough to be comparable to Planck area, then it is meaningless to talk about equilibrium IH as it would tremendously Hawking radiate and will be in a very dynamical situation -- the whole framework becomes inappropriate. Also, the number of punctures $N$ has to be large enough for suitably defining the area operator for a surface\cite{area1,area2,op1}.
\subsection{Equations of state}
Two important by products of this entropy calculations viz. eq.(\ref{lmro}) and eq.(\ref{siglm}) can be regarded as the equilibrium equations of state for the QIH relating $k,N,\lm$ and $\sigma$, and are of utmost relevance in the present context. 
Using $k=A_{cl}/4\pi\g\lp^2$, eq.(\ref{lmro}) can be rewritten as
\ba
A_{cl}=\tilde\alpha(\lm,\g) N\lp^2\label{ansatz1}
\ea
where $\tilde\alpha(\lm,\g)=4\pi\g\(1+\f{2}{\lm}+\f{4}{\lm(\sqrt 3\lm+2)}\)$.  Since $\lm$ is a function of $k/N$, which is transparent from eq.(\ref{lmro}) and eq.(\ref{siglm}), the above equation, although looks like the BM ansatz, but is actually not. Moreover, $\g$ is also unknown. Hence, fixing $\lm$ and $\g$ to particular values will get us to the desired result which I am going to do in the next sections.

\subsection{Fixing $\tilde\alpha$}

\subsubsection{Local vs Asymptotic Views}
The term $N\s$ in the entropy in eq.(\ref{lent}) originates from the consideration of $N$ as a macroscopic variable alongside $k$. This also gives rise to a term proportional to $\d N$ in the first law of thermodynamics associated with the QIH, which is only relevant to local observers close to the IH\cite{gp}. But there is no such term in the first law of thermodynamics associated with the IH derived from the classical theory\cite{ih3}. The only explanation which can consistently connect the two scenarios is that the quantity $\sigma$ must vanish in the classical limit and an asymptotic observer will {\it not} see the term $N\s$ in the expression for the entropy or equivalently the term proportional to $\d N$ is the first law.

\subsubsection{Fixing $\lm$ and $\g$}
Since the observation from the perspective of an asymptotic observer is only relevant in the context of black hole radiation, one must set $\s=0$ which is the only choice of $\s$ compatible with the first law of IH in the classical theory\cite{ih3} with is devoid of any term proportional to $\d N$, as mentioned earlier and explained in full detail in \cite{me1}. A graphical analysis of the eq.(\ref{siglm}) shows that $\s=0$ implies $\lm=1.2$ \cite{me1}. Hence, expression for the entropy in eq.(\ref{lent}) reduces to
$S={1.2 A_{cl}}/{8\pi\g\lp^2}$. Finally $\g=1.2/2\pi$ is chosen so as to yield  the area law\cite{bhal} $S=A_{cl}/4\lp^2$ as should be observed by the asymptotic observer.

\subsubsection{Value of $\tilde\alpha$}
Now, having fixed the values of $\lm$ and $\g$, one can calculate the value of $\tilde\alpha$ to be approximately $8.362$. Hence, it can be concluded that the BM ansatz is actually derivable in the LQG framework and is given by 
\ba
A_{cl}&=&\tilde\alpha N\lp^2\label{an.}
\ea
where $\tilde\alpha\simeq 8.362$. It should be mentioned that one can numerically calculate a refined and improved value of $\tilde\alpha$ by studying eq.(\ref{lmro}) in its original discrete form in \cite{me1}, which is expected to differ only slightly from this approximate one.

\subsection{The quantum number $N$}
It is to be noted that there is a crucial difference between what has been derived here in eq.(\ref{an.}) and the assumption made by BM given by eq.(\ref{an}). While BM have called eq.(\ref{an}) to be the `area spectrum', eq.(\ref{an.}) is the macroscopic area obtained from the equilibrium equations of state of the QIH by taking the classical limit. It is interesting to see that over and above the  quantum area spectrum of the QIH given by eq.(\ref{area}), there is a manifest discreteness of the area of the black hole horizon at the macroscopic level. The most significant issue regarding the above result is the appearance of $N$, the number of punctures, at the macroscopic level which is the sole reason for the manifest discreteness of the classical area of the black hole. If we trace the origin of the punctures or equivalently the variable $N$, it appears to be a mere mathematical structure needed to properly define the action of the area operator on a surface divided into $N$ pieces, each piece being small enough to contain only one piercing edge of the bulk spin network spanning the ambient spatial bulk quantum geometry in the limit $N\to\infty$\cite{area1,area2,op1}. This is just the limit,  we generally take, of a Riemann sum to define an integral. But, eq.(\ref{an.}) lifts the status of $N$ from just a mathematical structure to a physically consequential variable affecting the classical physics of black holes.

\section{Conclusion} 
I shall conclude with some remarks on the radiation spectrum one can expect to observe while a black hole ( depicted as IH here) decays from one equilibrium$(A_{cl1}=\tilde\alpha N\lp^2)$, to the immediate next one$(A_{cl2}=\tilde\alpha (N-1)\lp^2)$ through a dynamical phase. The effective realization of the BM ansatz imply that LQG predicts discrete radiation spectrum from black holes. Added to this, a closer observation shows that there will be broadening of the discrete lines. Since, a quantitative estimation of this effect is beyond the scope of this paper and needs further research, I shall provide a qualitative discussion on this particular issue and how it is already explicitly manifest from what has been discussed in this paper. It goes as follows. 

  Fixing $\s=0$ and hence, $\lm=\lm_0=1.2$, yields that the most probable spin distribution is given by $\sj[N]=N(2j+1)\exp\left[-\lm_0\j\right]$.
Clearly, the equation is parametrized by $N$. This shows that over and above the emission of a single frequency due to a jump from $N$ to $N-1$, there will be subdominant contributions due to the microscopic transitions between $\sj[N]$ and $\sj[N-1]$ which is expected to give rise to a broadening of the line spectrum. To check whether this broadening will be thermal in nature, is a task to be performed. It is quite possible that the dynamics of the punctures will play the crucial role and one may has to go beyond the kinematical structure of QIH.  

 The discrete line spectrum along with the line broadening is exactly what was also predicted by BM in \cite{bm}. That the line broadening results from the degeneracy causing the black hole entropy is explicitly manifest in this black hole description in the LQG framework, although not quantitatively estimated here. Consequently, the effective realization of BM ansatz within the LQG  framework, in the limit which is relevant for black hole thermodynamics,  puts the prediction of discrete spectrum for radiation of macroscopic black holes  on a firm ground of an underlying quantum theory of gravity. It will be interesting to see  whether a quantitative estimate of the line broadening in this LQG framework also matches the results of \cite{bm} in detail. Needless to say, it will make the case of \cite{bm} even stronger, albeit within the realms of LQG.

\vspace{1cm}

{\bf Acknowledgment :}  Initial stages of this research work was funded by the Department of Atomic Energy of the Government of India. The later stages of the work has been funded by the DST-Max-Planck partner group on Cosmology and Gravity in India.

\end{document}